\documentclass{egpubl}
\usepackage{EGUKCGVC2024}
 
\Poster                 
\WsPaper           

\usepackage[T1]{fontenc}
\usepackage{dfadobe}  

\usepackage{cite}  

\BibtexOrBiblatex
\electronicVersion
\PrintedOrElectronic

\ifpdf \usepackage[pdftex]{graphicx} \pdfcompresslevel=9
\else \usepackage[dvips]{graphicx} \fi

\usepackage{egweblnk} 

\usepackage[pass]{geometry}
\title{Towards a Generative AI Design Dialogue}
\author[A.E.Owen \& J.C.Roberts]{
\parbox{\textwidth}{\centering
Aron E. Owen\thanks{email: aron.e.owen@bangor.ac.uk}\orcid{0000-0001-5660-5867} and Jonathan C. Roberts\thanks{email: j.c.roberts@bangor.ac.uk}\orcid{0000-0001-7718-3181}
}\\
{\parbox{\textwidth}{\centering Bangor University, UK}
       }
}

\renewcommand{\backref}[1]{}
\renewcommand{\backrefalt}[4]{}
\begin{document}
\maketitle
\begin{abstract}
    Traditional visualisation designers often start with sketches before implementation. With generative AI, these sketches can be turned into AI-generated visualisations using specific prompts. However, guiding AI to create compelling visuals can be challenging. We propose a new design process where designers verbalise their thoughts during work, later converting these narratives into AI prompts. This approach helps AI generate accurate visuals and assists designers in refining their concepts, enhancing the overall design process. Blending human creativity with AI capabilities enables rapid iteration, leading to higher quality and more innovative visualisations, making design more accessible and efficient.
\ccsdesc[500]{Applied computing~Visualization}
\ccsdesc[300]{Computing methodologies~Artificial intelligence}
\printccsdesc   
\end{abstract}  
\section{Introduction}
We introduce the Design Dialogue Framework (DDF), a method inspired by a master blacksmith’s craftsmanship. DDF uses generative AI to transform initial sketches into polished visual designs, such as refining raw materials into an artefact. Traditional sketching captures ideas but often lacks the precision needed for detailed visualisations. DDF enhances this process with six steps: Define, Sketch, Describe, Engineer, Generate, and Evaluate, combining sketch-based ideation with generative AI to boost creativity and quality.
Generative AI, through prompt engineering, offers a structured approach to refining visualisation ideas. By crafting precise prompts, designers guide AI to produce visualisations that meet specific criteria, leading to innovative results.
This paper explores how generative AI can enhance visualisation design. DDF empowers designers to turn raw ideas into refined visualisations, unlocking new levels of creativity and inspiration.
\begin{figure*}[t]
  \centering
  \includegraphics[width=1\linewidth]{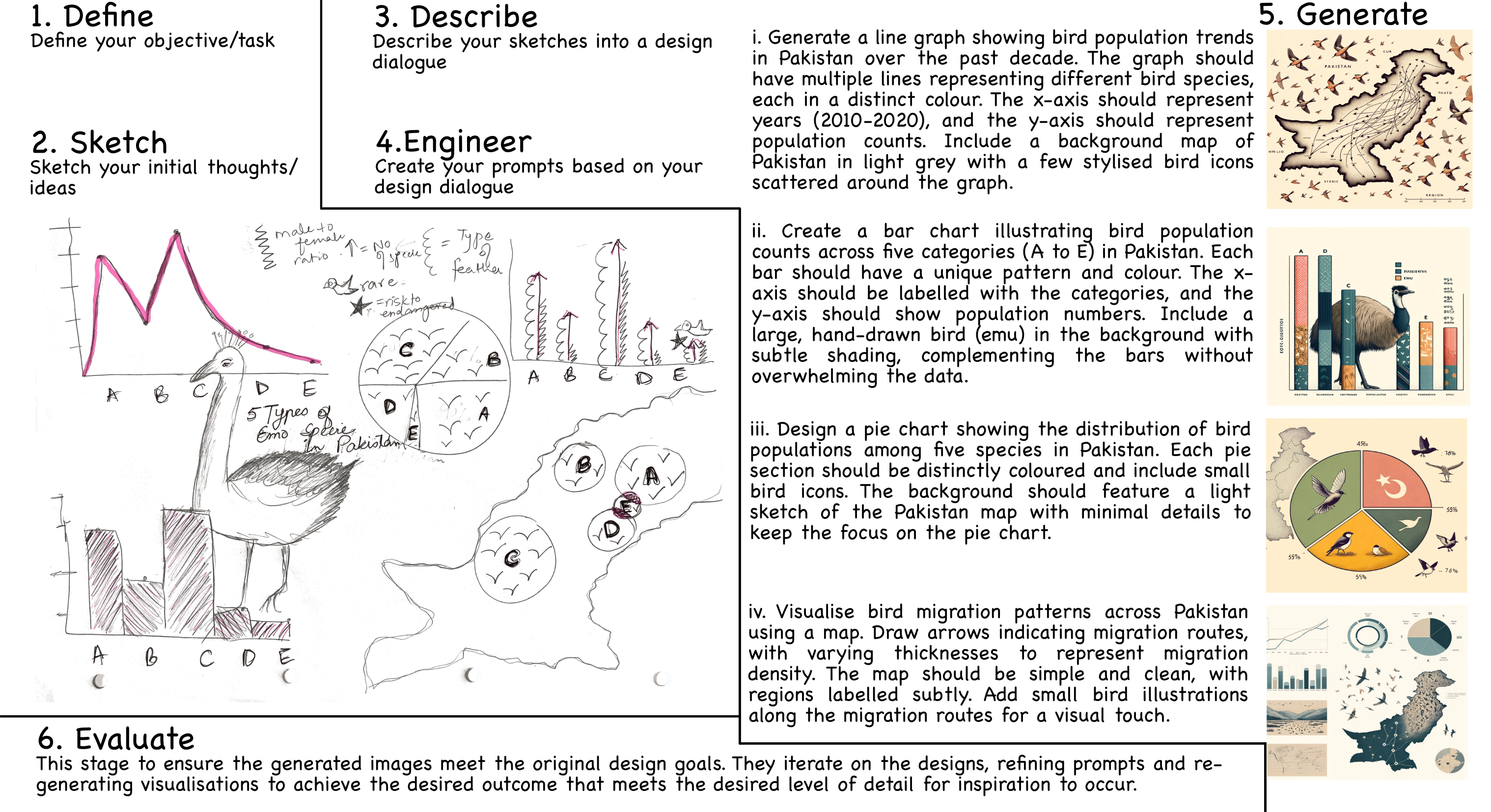}
  \caption{\label{fig:newFig}
           This image effectively demonstrates the DDF, an iterative visualisation methodology encompassing six steps: Define, Sketch, Describe, Engineer, Generate, and Evaluate. Embarking with defining the objective and sketching initial ideas, the process moves on to describing sketches in detail and engineering prompts for generative AI. The generated step delivers refined visualisations using bird population data and migration patterns in Pakistan. Ultimately, the evaluation step ensures the visualisation design goals, fostering iterative refinements and amplifying inspiration and creativity in the visualisation design process, are achieved.
           \vspace{-2mm}
}
\vspace{-2mm}
\end{figure*}
\section{Background and Related Work}
The Oxford English Dictionary states inspiration is ``A breathing in or infusion of some idea, purpose, etc. into the mind;  the suggestion, awakening, or creation of some feeling or impulse, esp. of an exalted kind''. Furthermore, it encourages us that inspiration requires us to think deeply, imagine concepts and draw on our previous feelings and experiences. Turning to existing literature and other frameworks helps us understand that inspiration requires thinking deeply, internalising concepts, and drawing on previous thoughts. Considering these ideas further, and from the literature~\cite{Wallas1926art, thrash2003inspiration, thrash2004inspiration, BertrandETALPaleoInspiration2018, Bederson2003craft, ShneidermanCreatingCreativity2000, owen2023visdice} people can be categorised as \textit{inspirational} (they receive moments of inspiration), \textit{structuralists} proceed in an orderly manner and \textit{situationalists}, relying on social and personal contacts. 
Nevertheless, regardless of the catalyst, it's essential for people to shift their perspectives to find renewed inspiration ~\cite{KnollHorton2011} and acknowledge its significance. The rapid advancement of LLMs and generative AI can quickly render some literature outdated. However, it remains crucial to recognise the contributions of all literature and build upon a strong foundation.

\noindent\textbf{Generative AI}: Many exciting research projects exist, such as integrating traditional art colours through generative AI, which correlates modern graphic and interactive design~\cite{Li2023, Zhang2023}. Complementing this, the work by~\cite{Tikfan2023} assesses generative AI's ability to inform HCI technology's future. Enhancing narrative immersion through AI-driven tactile experiences adds a tangible dimension to storytelling~\cite{Chiang2023}. Integrating AI with storytelling and cultural motifs paves the way for innovative, future-oriented narratives in design~\cite{Fu2023}.
Some clarification is needed on whether generative AI tools will be used in product ideation/inspiration, guiding us towards a future where generative AI will aid designers~\cite{Ye2023}. In particular, generative AI can significantly help visualisation storytelling, and  as skill requirements shift from creative to technical we can expand the horizons of design thinking~\cite{Zhu2023}. For instance, PlotThread showcases how reinforcement learning can enhance the narrative coherence and aesthetic appeal of visual stories, highlighting practical applications of AI in refining the design and communication of narratives~\cite{Tang2023}.

\noindent\textbf{Prompt Engineering}: Like many other design aspects, prompt engineering is an emerging skill, presenting a promising step forward for the future of design dialogues and interactive storytelling~\cite{Harmon2023}. Investigating interactive systems like PromptMagician, which refine prompts for text-to-image generation, highlights the potential of user-centric design~\cite{feng2024}. The development of automatic systems for prompt engineering demonstrates the advanced systems used for improving LLM's efficiency in its tasks~\cite{WANG2023}. 
Frameworks like CLEAR (Concise, Logical, Explicit, Adaptive, and Reflective) underscore the importance of structured methodologies in effectively interacting with AI models~\cite{framework2023}. GitHub Copilot excels in solving programming problems but is only possible with accurate use~\cite{denny2023}. Research on visual prompt engineering for Artificial General Intelligence (AGI) highlights the importance of incorporating visual cues to improve model understanding and performance~\cite{Strobelt2023}. There is a clear need for prompt engineering in visualisation~\cite{SHORT2023}.
\section{Design Dialogue Framework}
The DDF enhances visualisation design by merging traditional sketching with generative AI. Inspired by a blacksmith’s craft, it guides designers through six stages: Define, Sketch, Describe, Engineer, Generate, and Evaluate. The goal is to turn initial ideas into high-quality visual designs, with AI boosting creativity and efficiency.
Designers first define the concept and sketch ideas quickly. Next, they describe the best sketches in detail, providing a blueprint for AI-generated visuals. Structured prompts guide AI in generating diverse visualisations, which are then evaluated and refined to meet design goals. This iterative process bridges traditional and AI-driven design, making the workflow more accessible and innovative and leading to compelling visual results.

\section{Discussion and Conclusion}
This paper underscores the potential of the Design Dialogue Framework (DDF) to revolutionise visualisation design. By integrating traditional sketching techniques with the power of generative AI, the DDF has shown promising results in a usage scenario in Fig \ref{fig:newFig} on visualising bird population data and migration patterns in Pakistan. The framework’s structured and iterative approach fosters creativity and enables rapid refinement, simplifying the exploration and development of sophisticated visualisations. While the DDF highlights the powerful synergy between AI and human creativity, it also brings to light significant challenges, such as ensuring data quality and maintaining a balance between automation and human oversight. Future research should focus on refining the DDF further, exploring the intricacies of human-AI collaboration, and expanding the framework’s applicability across diverse fields. These efforts will help unlock the full potential of the DDF, making visualisation design more accessible, practical, and innovative and ultimately contributing to the advancement of the field.
\bibliographystyle{eg-alpha-doi} 
\bibliography{main}       

\end{document}